\documentstyle[twoside,fleqn,espcrc2,psfig]{article}
\title{
Muon and Muon Neutrino Fluxes from Atmospheric Charm
}
\author
{L. Pasquali\address{
Department of Physics and Astronomy, University of Iowa,
Iowa City, IA 52242 USA}, M. H. Reno$^{\rm a,}$\address{
CERN-TH, CH-1211 Geneva 23, Switzerland}%
\thanks{Talk presented
by M. H. Reno. Work supported in part by NSF Grant No.
PHY-9507688 and D.O.E. Contract No. DE-FG02-95ER40906.} 
and   I. Sarcevic\address{
Department of Physics, University of Arizona,
Tucson, AZ 85721 USA}
}
\begin{document}

\begin{abstract}
The charm contribution to the atmospheric fluxes of muons and muon
neutrinos may be enhanced by as much as a factor of 10 when one
includes the contributions of $D\rightarrow\pi ,K\rightarrow$leptons
and folds in uncertainties in the charm cross section and energy
distribution. In the energy range considered here, from 100 GeV to
10 TeV, the charm contribution is small compared to the
conventional flux of muons and muon neutrinos.
\end{abstract}

\maketitle

The fluxes of leptons from the decays of pions and muons produced by
cosmic ray interactions in the atmosphere are known to within
approximately $\pm 20$\%\cite{group}
at energies $\sim 1$ GeV. At cosmic ray energies greater than
a few GeV, charm-anticharm pairs can be produced. The semileptonic
decays  of charmed mesons and baryons which emerge from the
cosmic ray interactions with air are additional contributions to the
atmospheric lepton fluxes. Here, we present the charm contribution to the
atmospheric lepton fluxes in the energy range of 100 GeV to
10 TeV. We evaluate the theoretical uncertainties associated with 
charm
production
and decay. We show that the lepton flux from charm decays
has a factor of approximately
ten uncertainty, but the uncertainty has little implication for the
measured atmospheric muon and muon neutrino fluxes in this energy
range.

Atmospheric leptons from pion and kaon decays are called ``conventional''
leptons. Leptons from charm decays contribute to the ``prompt'' flux.
Recently, a new calculation of the prompt lepton fluxes by
Thunman, Ingelman and Gondolo (TIG)\cite{tig} has appeared in
the literature. Using a PYTHIA\cite{pythia}
based Monte Carlo, they have evaluated
the contributions of semileptonic decays of charmed particles to
the leptonic fluxes. One aspect of charm decays not included 
in other calculations is
the decay chain of charm$\rightarrow \pi,K\rightarrow$leptons, called
here ``secondary'' contributions. Each charm decay has an associated
pion and/or kaon in the final state, which itself decays leptonically.

The calculation of prompt, conventional and secondary lepton fluxes
can be done approximately using the $Z$-moment form of the
cascade equations. Details of approximate solutions can be found
in, for example, Ref. \cite{lipari}. The general form for the flux
of particles of type $j$, as a function of energy and
slant depth in the atmosphere, $X$, is
\begin{equation}
{d\phi_j\over dX}  = -{\phi_j\over \lambda_j}
-{\phi_j\over \lambda_j^{(dec)}}+\sum_k S(k\rightarrow j) \nonumber
\end{equation}
where
\begin{equation}
S(k\rightarrow j)  = \langle N_j\rangle\int_E^\infty
dE_k {\phi_k(E_k,X)\over \lambda_k(E_k)}{dn_{k\rightarrow j}\over
dE}
\end{equation}
The quantities $\lambda_j$ and $\lambda_j^{(dec)}$ are the interaction and
decay lengths, $\langle N_j\rangle$ is the average particle multiplicity of
type $j$, and $dn_{k\rightarrow j}/dE$,
describes the energy distribution
of particle $j$ given its production by particle $k$ with energy $E_k$.
$Z$-moments are defined by
\begin{equation}
S(k\rightarrow j)
\equiv {\phi_k(E,X)\over \lambda_k(E,X)} Z_{kj}(E) \ .
\end{equation}
The $Z$-moments $Z_{kj}$ depend on particle type and energy.

The starting point for the solution to the cascade equations is the
cosmic ray flux, which we take as all protons with an energy behavior
$\phi_p\sim E^{-2.7}-E^{-3}$, as in TIG.
Given the cosmic ray flux, and the interaction and decay moments of TIG in
Ref. \cite{tig}, we can evaluate the conventional and prompt fluxes.
Since our interest in is the ``low energy'' regime, 100 GeV$<E<10$ TeV,
we include only $c=D^0,\ \bar{D}^0,\ D^\pm$ in our charm contribution.
TIG have shown that these are dominant in this energy range.
The new feature here is to include secondary decays:
we have additionally, for example, $D\rightarrow\pi\rightarrow \mu$,
so we need the decay moment $Z_{D\pi}$.
As a first approximation, we rescale the decay moments for $D$'s into
neutrinos to account for hadronic branching fractions and multiplicities.
The details of the $D$ decay-moment inputs can be found in
Ref. \cite{prs}.

Our results for the fluxes of muons (particles plus
antiparticles),
including secondary decays, are shown in Fig. 1 for the vertical
direction. The solid line indicated by P (prompt) include only $D^0$,
$\bar{D}^0$ and $D^\pm$. The dashed lines are the TIG parameterization of
their Monte Carlo results \cite{tig}, which have significant $\Lambda_c$ and
$D_s^\pm$ contributions at high energies. At low energies, the $D$'s
dominate. The muon neutrino fluxes are similar. The prompt flux of
muon neutrinos equal the prompt flux of muons. The conventional 
muon neutrino flux
is approximately a factor of 10 suppressed
relative to the muon flux. The secondary muon neutrino flux is about a factor
of five suppressed relative to the secondary muon flux.

\begin{figure}[htb]
\psfig{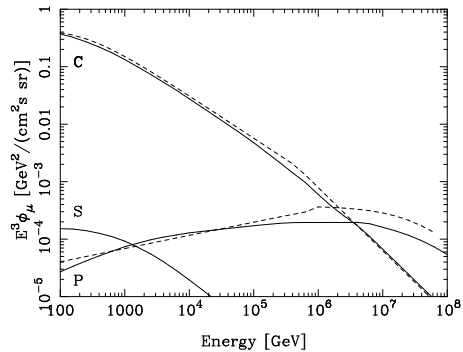}
\caption{Conventional (C), prompt (P) and secondary (S) atmospheric
muon plus antimuon
flux in the vertical direction. The dashed lines are the approximate
formulae of Ref. [2].}
\end{figure}

As can be seen from the figure, the secondary flux is approximately
three orders of magnitude below the conventional one. As one goes to
angles off the vertical, both the secondary and conventional fluxes
increase, but in a constant ratio, while the prompt flux remains
constant as a function of angle.

Uncertainties in the calculation of the charm cross section
and energy distribution affect predictions of both the prompt and
secondary fluxes. We estimate that the
cross section data \cite{mangano} can accommodate an additional factor of two
in the theoretical prediction of TIG \cite{tig}.
The energy distribution in the charm production
cross section is another uncertainty. The energy distribution is usually
written in terms of the charm particle energy divided by the incident
nucleon (beam) energy:
$x=E_c/E_b$, and in the scaling approximation,
$d\sigma/dx \sim (n+1)(1-x)^n$. 
Next-to-leading order (NLO) perturbation theory, when fit to the
$(1-x)^n$ distribution, has $n\sim 6-9.5$ 
for $E_b=100-1000$ GeV, while
experimental measurements yield $n\sim 4.9-8.6$ in the same energy
range\cite{mangano}.  
By using experimental rather than theoretical values for $n$,
the charm flux can be enhanced by a factor of 1.5.

Finally, the charmed meson decay moment used for Fig. 1 is based on
a parton V-A formula. If we use phase space instead of the V-A formula,
the $Z$ decay moment is enhanced by a factor of 2.4.
Taken together, these enhancements  can increase the secondary
flux by a factor of 7, and the prompt flux by a factor of 3.

\begin{figure}
\psfig{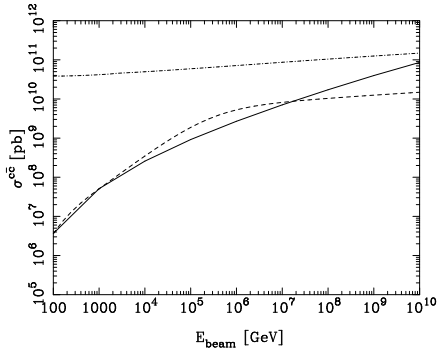}
\caption{The charm-anticharm cross section according to Eq. (4) (dashed
line). The solid line is the NLO charm cross section using
CTEQ3 parton distribution functions with $\mu=m_c=1.3$ GeV 
and the dot-dashed line is 
$\sigma_{pp}$.}
\end{figure}

\begin{figure}
\psfig{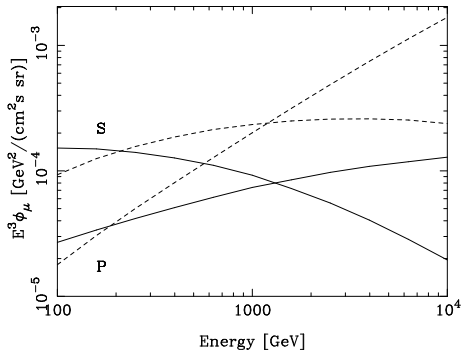}
\caption{A comparison of the prompt (P) and secondary (S) muon fluxes
using TIG parameters (solid line) and the fluxes computed using
the cross section of Eq. (4) and a scaling energy behavior with
$n=4$.}
\end{figure}

To estimate the effect of the high energy cross section on the
low energy flux, we use a cross section which becomes $0.1\sigma_{pp}$
at high energies
as suggested by Zas et al.\cite{zas}, 
where $\sigma_{pp}$ is the total $pp$ cross section\cite{pdg}:
\begin{equation}
\sigma^{c\bar{c}} = {\sigma^{LE}\times 0.1\sigma_{pp}
\over \sigma^{LE}+0.1\sigma_{pp}}
\end{equation}
where $\sigma^{LE}$,
for $E=100-1000$
GeV, is the next-to-leading order (NLO) charm pair production
cross section, evaluated at factorization and renormalization scales 
$\mu$ equal to $m_c=1.3$
GeV, using the CTEQ3 parton distribution functions\cite{CTEQ}.
We use a power law
extrapolation for $E>1$ TeV. The cross section of Eq. (4) is represented
by the dashed line in Fig. 2.
As an extreme, we take $n=4$ in $d\sigma/dx\sim (1-x)^n$, barely consistent
with the low energy data. Our results for the muon plus antimuon fluxes
are shown in Fig. 3.

In conclusion,
there is a factor of about ten uncertainty in the prediction for
the flux of muons from the decay of charm in the energy range
100 GeV-10 TeV. 
Relative to the prompt muon flux, the secondary decay contribution
is significant, however, relative to the conventional flux, it is not.
These conclusions also apply to the muon neutrino fluxes from charm decay.
Because the conventional electron neutrino flux is small, the
prompt flux is more important at 10 TeV than in the
muon neutrino and muon case. This is a topic under further investigation
by the present authors.

\end{document}